\documentclass[%
 aip,
 amsmath,amssymb,
reprint,%
]{revtex4-2}

\usepackage{graphicx}
\usepackage{dcolumn}
\usepackage{bm}
\usepackage{hyperref}
\usepackage[utf8]{inputenc}
\usepackage[T1]{fontenc}
\usepackage{mathptmx}
\usepackage{etoolbox}
\usepackage{enumitem}
\usepackage{xcolor}
\usepackage{easyReview}
\usepackage{amsmath}

\makeatletter
\def\@email#1#2{%
 \endgroup
 \patchcmd{\titleblock@produce}
  {\frontmatter@RRAPformat}
  {\frontmatter@RRAPformat{\produce@RRAP{*#1\href{mailto:#2}{#2}}}\frontmatter@RRAPformat}
  {}{}
}%
\makeatother
\begin{document}

\preprint{AIP/123-QED}

\title[Enhancing model identification with SINDy via nullcline reconstruction]{Enhancing model identification with SINDy via nullcline reconstruction}
\author{Bartosz Prokop}
\author{Nikita Frolov}%
\author{Lendert Gelens}
\email{Bartosz.Prokop@kuleuven.be, Lendert.Gelens@kuleuven.be}
\affiliation{ 
Dynamics in Biological Systems, KU Leuven, 3000 Leuven, Belgium 
}%


\date{\today}

\begin{abstract}
Many dynamical systems exhibit oscillatory behavior that can be modeled with differential equations. Recently, these equations have increasingly been derived through data-driven methods, including the transparent technique known as Sparse Identification of Nonlinear Dynamics (SINDy). This paper illustrates the importance of accurately determining the system's limit cycle position in phase space for identifying sparse and effective models.
We introduce a method for identifying the limit cycle position and the system's nullclines by applying SINDy to datasets adjusted with various offsets. This approach is evaluated using three criteria: model complexity, coefficient of determination, and generalization error.
We applied this method to several models: the oscillatory FitzHugh-Nagumo model, a more complex model consisting of two coupled cubic differential equations with a single stable state, and a multistable model of glycolytic oscillations. Our results confirm that incorporating detailed information about the limit cycle in phase space enhances the accuracy of model identification in oscillatory systems.
\end{abstract}

\maketitle

\begin{quotation}
Data-driven model identification methods have reshaped how scientists infer models of dynamical systems from data. These techniques enable developing predictive models to explore these systems, requiring minimal prior knowledge. Specifically, when white-box methods like the SINDy approach are used, they allow for the derivation of mathematical models expressed as differential equations. However, despite their clear benefits, these methods are rarely utilized to extract explicit models directly from experimental data. The main reason for this is the sensitivity of such methods to low (spatio-)temporal resolution or high noise levels in the data. Another significant obstacle is the absence of crucial additional information about the system, which could markedly enhance the accuracy of model identification or discovery. Such information includes the accurate position of the data within the system’s phase space — typically obscured by the measurement of intermediate variables or data cleaning processes. In this work, we suggest that identifying the correct data positioning and other relevant information can be identified by shifting the system's attractor within the phase space and analyzing the results obtained with SINDy. Our experiments with the Fitzhugh-Nagumo equation and more complex oscillatory systems demonstrate that with adequate resolution, it is feasible to recover such additional information, thereby enriching our understanding of the underlying system and enhancing the effectiveness of SINDy and similar model identification techniques.
\end{quotation}

\section{Introduction}
\label{sec:introduction}
Many natural processes are characterized by periodic changes or oscillations, such as the early embryonic cell cycle \cite{Murray1989}, circadian rhythms \cite{Patke2019}, cardiac rhythms \cite{Mackey1977,Glass2006}, and population dynamics \cite{Volterra1928}.
Understanding and describing the typically nonlinear mechanisms of these oscillations is crucial.
Therefore, researchers have developed mathematical models using differential equations to generally describe oscillatory behavior.
These models are often derived by analyzing experimentally collected data, proposing possible equations based on scientific knowledge and established model design principles \cite{Newton1729,Novak2008}, and identifying the best-fitting and most generalizable equations.

However, the increasing generation of large and complex datasets, commonly referred to as big data, across various natural sciences has made model derivation from first principles and scientific intuition more challenging \cite{Miller1956, Leonelli2019, Murari2020}.
Over recent decades, numerous data-driven methods have emerged to derive interpretable mathematical equations directly from measured (spatio)temporal data.
These include the Nonlinear Autoregressive Moving Average Model with Exogenous Inputs (NARMAX) \cite{billings2013nonlinear}, Symbolic Regression \cite{Schmidt2009}, and Sparse Identification of Nonlinear Dynamics (SINDy) \cite{Brunton2016}.

SINDy has been effectively applied to synthetic data across fields such as engineering \cite{Reinbold2021}, physics \cite{Ermolaev2022}, chemistry \cite{Hoffmann2019}, and biology \cite{Mangan2016}.
Yet, it faces limitations with low data quality and observation time \cite{Thaler2019,Fasel2022,Hirsh2022,Prokop2023}.
In biologically-motivated systems and initial experimental studies with high-resolution data on mostly transient predator-prey dynamics \cite{Horrocks2020,Liang2022,Brummer2023}, SINDy has shown promising results.
However, in environments characterized by low data availability or high noise, it struggles to identify interpretable models, as demonstrated in our recent work \cite{Prokop2024}.

A significant challenge is the absence of prior information about the dynamics of experimental systems. Access to such information, such as system stability \cite{Kolter2019} or vector field constraints \cite{Ahmadi2023}, can greatly enhance model identification by constraining the optimization space.
For oscillatory systems, relevant knowledge might include details about the limit cycle attractor in phase space, such as the location of fixed points or the system’s nullclines.
However, this information is often lost during measurements (e.g., through indirect microscopy imaging) or subsequent data cleaning processes (e.g., detrending), leading to the identification of non-sparse models.

In this study, we demonstrate how applying SINDy to synthetic data from the well-established Fitzhugh-Nagumo (FHN) oscillator model can pinpoint the precise position of the limit cycle in phase space and provide insights into the shape of the nullclines.
We assess the efficacy of our approach by evaluating the identified models based on their complexity (or sparsity), the coefficient of determination (R$^2$), and the proximity to the attractor ($\delta$).
Additionally, we illustrate that our method also works on more complex models involving two coupled cubic differential equations.
Using such additional information on fixed points and nullcline structure, we enhance the accuracy of model identification in oscillatory systems.

\section{Sparse Identification of Nonlinear Dynamics (SINDy)}

The fundamental principle of SINDy is based on the premise that a dynamical system can be succinctly represented by a sparse differential equation. The model assumes that the temporal dynamics of the system, expressed through state variables $u$ and $v$ (denoted as $u_t, v_t$), can be captured by a linear combination of these variables, their interactions, and corresponding coefficients $\bm{\xi}$:

\begin{align}
	\begin{split}
		\left( \begin{array}{c}u_t\\v_t\end{array}\right) &= N\left[u, v, uv, u^2, v^2,\dots, \bm{\xi}\right] \\ &= \left( \begin{array}{c}\xi_{11} + \xi_{12} u + \xi_{13} v + \xi_{14} uv + \dots\\\xi_{21} + \xi_{22} u + \xi_{23} v + \xi_{24} uv + \dots\end{array}\right).
	\end{split}
\end{align}

At each time point, this model formulation leads to a linear system where the unknown coefficient vector $\bm{\xi}$ and the various term combinations form the term library matrix $\bm{\Theta}$:

\begin{align}
	\begin{split}
		\begin{pmatrix} \vline&\vline \\ u_t & v_t \\\vline &\vline \\\end{pmatrix} &= \begin{pmatrix}
			\vline & \vline & \vline & \vline & \vline & \vline \\1 & u & v & uv & u^2 & ... \\  \vline & \vline & \vline & \vline & \vline & \vline\\\end{pmatrix} \cdot \bm{\xi}=\bm{\Theta}\cdot\bm{\xi}.
	\end{split}
\end{align}

This creates an over-determined system, ideal for optimization aimed at determining the values of $\bm{\xi}$ using standard regression methods. In this study, we utilize a term library up to third order, incorporating terms such as $1, u, v, u^2, uv, v^2, u^3, u^2v, uv^2, v^3$. We employ ridge regression with sequential thresholding as initially introduced by Brunton \textit{et al.}\cite{Brunton2016}:

\begin{align}
	\label{eq:ridge_regression}
	\min_{\bm{\xi}}\frac{1}{2} \|\bm{X_t}- \bm{\Theta}\bm{\xi}\|^2_2 + \alpha \| \bm{\xi} \|_2 
\end{align}

Here, $\alpha$ is the sparsity-promoting factor that penalizes significant non-zero values in $\bm{\xi}$, set to $\alpha= 10^{-20}$ following our prior research \cite{Prokop2024}. Model sparsity is achieved by discarding coefficients $\left|\xi_{ij}\right|$ that fall below a predefined threshold $\xi_\text{thres}$, setting them to zero: $\left|\xi_{ij}\right|< \xi_\text{thres} \rightarrow \xi_{ij}=0$.

\section{\label{sec:methods} Impact of variable repositioning on SINDy model identification}

In studies applying SINDy to synthetic data, the limit cycle attractor is typically correctly positioned in the phase space. We investigated the effects of non-optimally positioned limit cycles on model identification.

\begin{figure}
\includegraphics[width=1\linewidth]{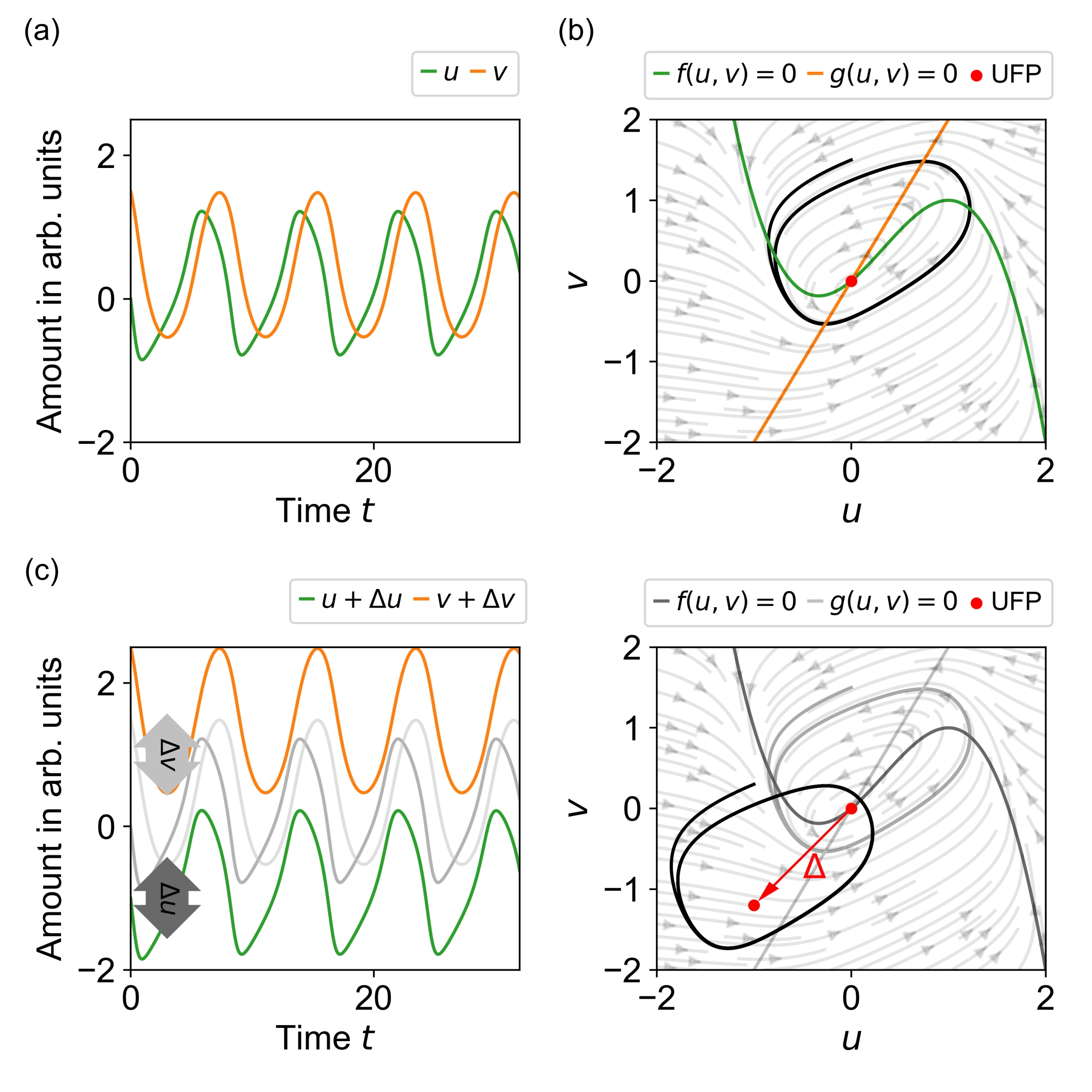}
\caption{\label{fig:fhn_ts} (a) Simulated time series of the FHN equation (Eq.~\ref{eq:fhn}), (b) the limit cycle with nullclines and an unstable fixed point (UFP). (c) Introducing an offset $\Delta = (\Delta u, \Delta v)$ repositions the time series and limit cycle in the phase space. Parameters: $a=0, b=0.5, c=1, d=1, \varepsilon=1$.}
\end{figure}

The FHN model, initially described by Fitzhugh \cite{FitzHugh1961} and Nagumo \cite{Nagumo1962}, is used here. Originally developed to model neuronal excitability, it has also been broadly applied across fields like cardiology, the cell cycle, and even in designing electronic circuits and all-optical spiking neurons \cite{Cebrian-Lacasa2024}.
We describe the system with:
\begin{align}
    \begin{split}
        \label{eq:fhn}
        u_t &= f(u,v) = -u^3 + cu^2 + du - v,\\ 
        v_t &= \varepsilon g(u,v) = \varepsilon(u - bv + a), 
    \end{split}
\end{align}
and its nullclines:
\begin{align}
    \begin{split}
        \label{eq:fhn_nullcline}
        0 &= f(u,v) = -u^3 + cu^2 + du - v,\\ 
        0 &= g(u,v) = (u - bv + a). 
    \end{split}
\end{align}
\begin{figure*}
	\includegraphics[width=1\linewidth]{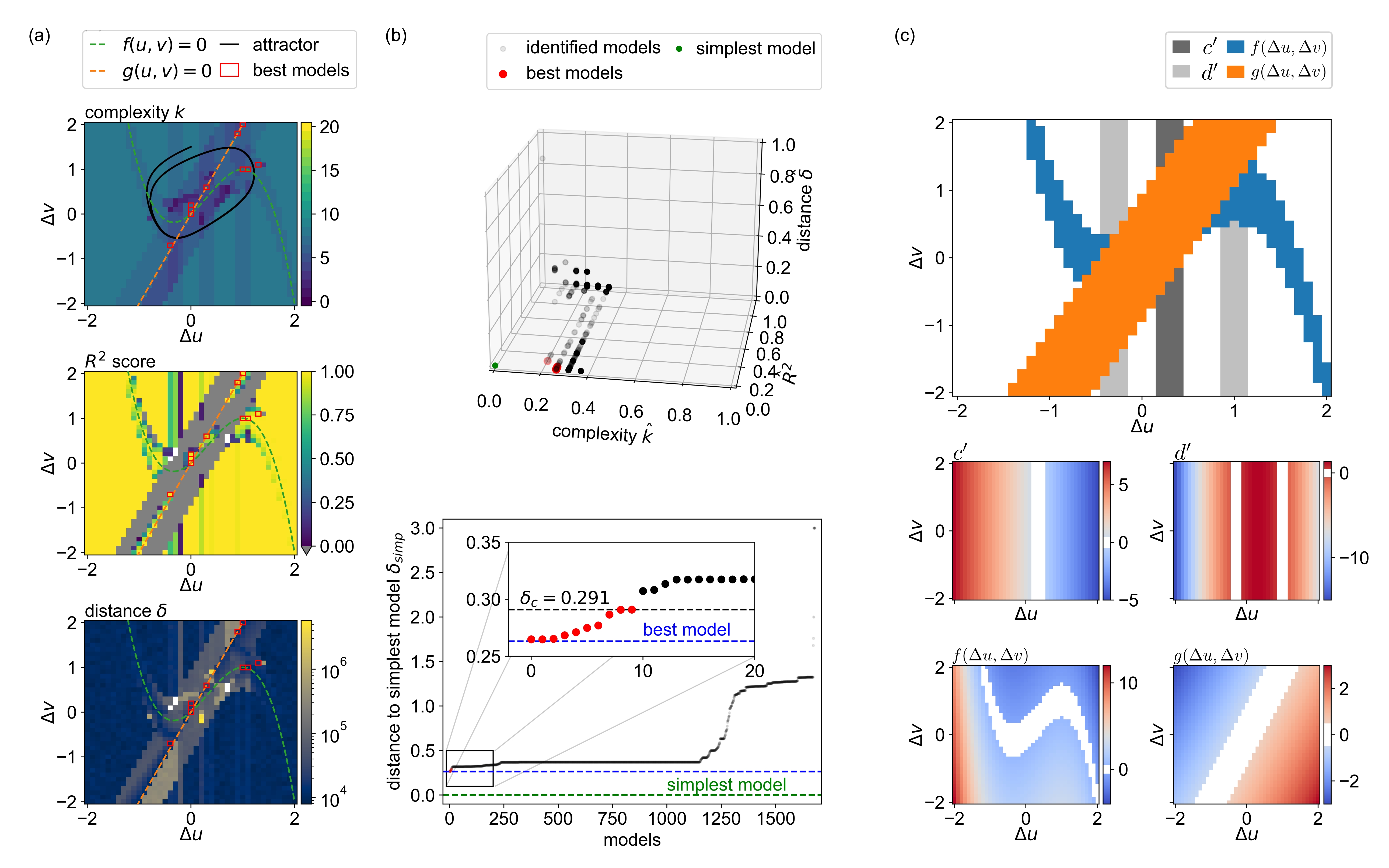}
	\caption{\label{fig:high_res} \textbf{Analysis of FHN model inference for different offsets} (a) Evaluation of complexity $k$, $R^2$ score, and distance $\delta$ for $dt=0.01$, highlighting top-ranked models with visible nullcline structures. (b) The models are prioritized based on complexity $\hat{k}$, $\hat{R^2}$ score, and distance $\hat{\delta}$, combined in  $\delta_{\text{simp}}$, determining the best ten models through a cutoff $\delta_c$. (c) Examination of the patterns in the three criteria plots, influenced by the threshold (here $\xi_{\text{thres}}=0.49$). This reveals model coefficients for various offsets, showing vertical lines and structures corresponding to the nullclines.}
\end{figure*}
With parameters set as $a=0, b=0.5, c=1, d=1, \varepsilon=1$, we generate the time series, limit cycle, and nullclines. The v-nullcline is a straight line (orange), while the u-nullcline is S-shaped (green), crossing at an unstable fixed point (UFP) at $(u,v) = (0,0)$.
To explore the impact of limit cycle position on model identification, we maintain the cycle's form but alter the data by introducing offsets $\Delta u$ and $\Delta v$ (see Fig.~\ref{fig:fhn_ts}c,d). This modifies the original model to:
\begin{align}
\label{eq:fhn_offset}
    \begin{split}    
        u_t &= -u^3 + c'u^2 + d'u - v + f(\Delta u, \Delta v), \\
        v_t &= \varepsilon(u - bv) + \varepsilon g(\Delta u, \Delta v),
    \end{split}
\end{align}
where:
\begin{align}
\label{eq:coeff_offset}
    \begin{split}    
        c' &= -3 \Delta u + c, \\
        d' &= -3 (\Delta u)^2 + 2c\Delta u + d.
    \end{split}
\end{align}


\begin{figure*}
	\includegraphics[width=1\linewidth]{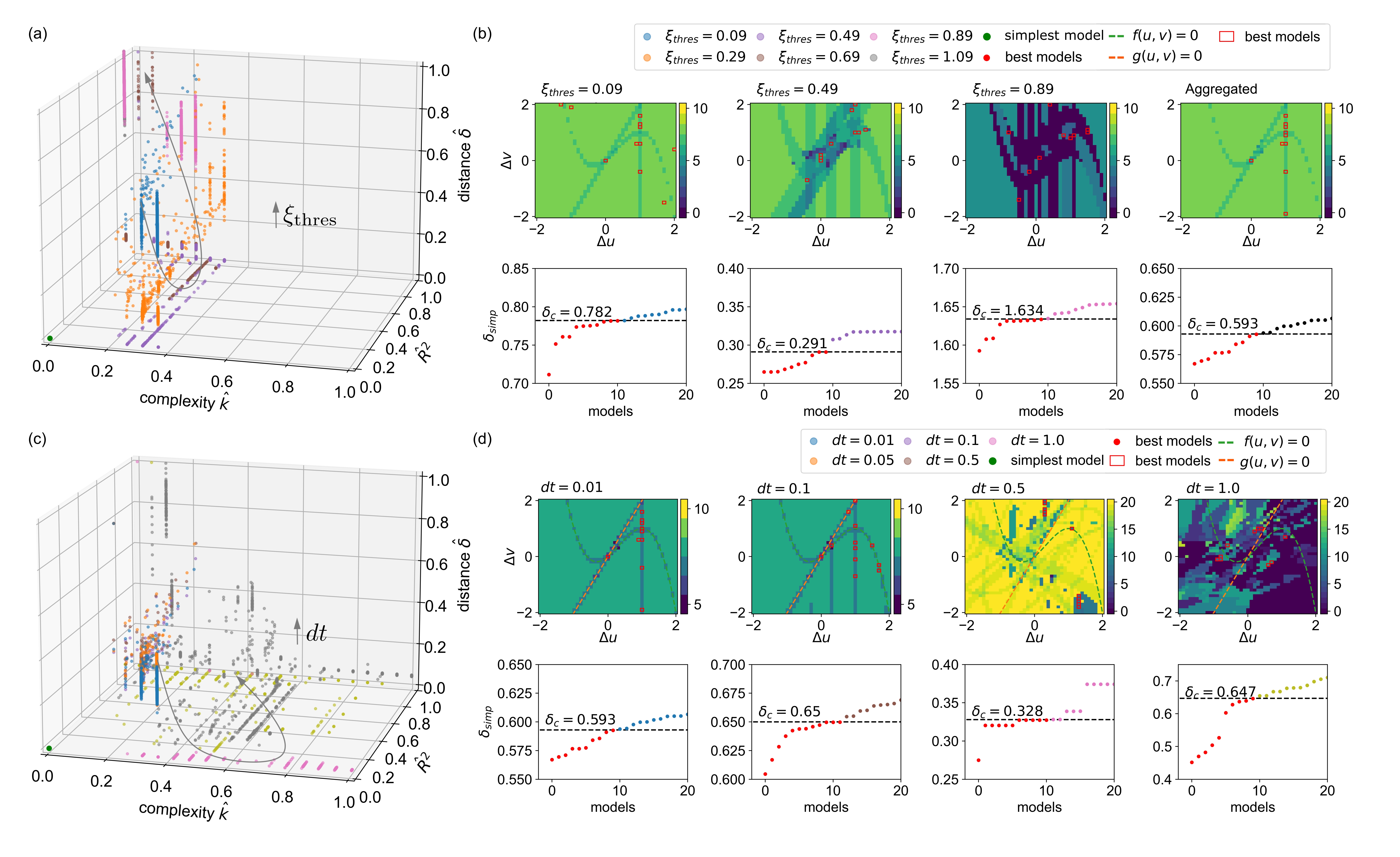}
	\caption{\label{fig:fhn_thresh_temp} \textbf{Influence of threshold $\mathbf{\xi_{\text{thres}}}$ and temporal resolution $\mathbf{dt}$ on model inference for the FHN model.}(a) Varying the threshold $ \xi_{\text{thres}}$ in SINDy improves the identification the closer the threshold lies to the minimal coefficient.For larger thresholds the distance to the attractor $\hat{\delta}$ and $R^2$ score increase. (b) Thresholds below the minimal coefficient ($\xi_{\text{thres}}<\xi_{\text{min}}$) reveal nullclines in complexity $k$ contour plots and improve ranking $\hat{\delta}_{\text{opt}}$ closer to 0. Without knowledge of $\xi_{\text{min}}$, aggregating results across thresholds can isolate models with the lowest $\hat{R}^2$ scores. (c) Reducing temporal resolution initially retains model accuracy with low $\hat{R}^2$ scores but increases complexity. Further reduction deteriorates model performance and its ability to capture the data. (d) Sufficient resolution $dt$ aids in visualizing nullcline structures and potential (unstable) fixed points through complexity $k$ contour plots, capturing the true fixed point. Lower resolutions increase complexity up to $k_{\text{max}}=20$, obscuring nullcline structures.}
\end{figure*}

Eq.~(\ref{eq:fhn_offset}) demonstrates that shifting the limit cycle in phase space results in altered model coefficients $c \rightarrow c'$, $d \rightarrow d'$, and introduces additional terms in each equation, represented by $f(\Delta u,\Delta v)$ and $g(\Delta u,\Delta v)$. These changes significantly affect SINDy model identification, potentially causing key coefficients to drop below the threshold $\xi_{\text{thres}}$, leading to either less sparse representations or failure in model identification.
We will evaluate these impacts by ranking models based on three criteria (see Appendix~\ref{app:criterion}):\\

\begin{description}[topsep=0pt, partopsep=0pt, itemsep=0pt, parsep=0pt]
\item[Complexity] Evaluates solution complexity or sparsity, preferring simpler models. Complexity $k$ is normalized against the maximum number of terms in the term library $k_{\text{max}}=20$.
\begin{equation}
    \hat{k}=\frac{k-k_{\text{simp}}}{k_{\text{max}}-k_{\text{simp}}}
\end{equation} 
with the simplest model complexity $k_{\text{simp}}=1$.\\
 
\item[Distance from attractor] Measures how closely a model's trajectory under varying initial conditions aligns with the original limit cycle. Smaller distances indicate a stronger attraction to the same attractor, enhancing model ranking. Distance $\delta$ is normalized by the maximum observed distance $\delta_{\text{max}}$,
\begin{equation}
	\hat{\delta}=\frac{\delta}{\delta_{\text{max}}}.
\end{equation} 

\item[$\mathbf{R^2}$ score] Represents the coefficient of determination, with 1 indicating a perfect match. The best models achieve the lowest inverted scores $\hat{R}^2=1-R^2$, coinciding with the simplest model where $\hat{k}=0$ and $\hat{\delta}=0$.
\end{description}

\section{Results\label{sec:results}}
\subsection{Varying attractor position to identify additional model information with SINDy}

\paragraph{Ranking models provides visual clues on the nullcline structure.}
\label{sec:FHN_org}

We simulated time series of the FHN equation (\ref{eq:fhn}) for four periods (as in our previous study \cite{Prokop2024}) with a time step of \(dt = 0.01\). The offset \((\Delta u, \Delta v)\) was varied between [-2,2] in increments of 0.1. Using these time series, we applied SINDy for every offset and plotted the complexity \(k\), the \(R^2\) score, and the distance \(\delta\) (see Fig.~\ref{fig:high_res}a). Visual inspection of the contour plots reveals structures that resemble the nullclines of the system. We overlay the nullclines \(f(\Delta u, \Delta v) = 0\) and \(g(\Delta u, \Delta v) = 0\) in green and orange, respectively, showing close alignment with the actual nullclines and intersecting at the correct (unstable) fixed point at \((\Delta u, \Delta v)=(0,0)\).\\

To determine the best model among the 1600 identified, we ranked them based on
the normalized measures of complexity $\hat{k}$, goodness of fit $\hat{R}^2$, and distance $\hat{\delta}$. We introduced a hyperparameter \(\delta_{\text{simp}}=\hat{k}+\hat{R}^2+\hat{\delta}\) to assess model quality (see Fig. ~\ref{fig:high_res}b). Although \(\delta_{\text{simp}}=0\) does not indicate the ground truth model (actual value approximately 0.2632, shown in Fig.~\ref{fig:high_res}b in blue), it helps evaluate models relative to the simplest configuration. Using \(\delta_{\text{simp}}\), we selected the top ten models (shown in Fig.~\ref{fig:high_res}b with red markers), which align closely with the nullclines, particularly near \((\Delta u, \Delta v)=(0,0)\), \((\Delta u, \Delta v)=(1,1)\), and \((\Delta u, \Delta v)=(1,2)\).\\

Further analysis of each model's coefficients along the nullclines showed that constant coefficients tend to vanish (Fig.~\ref{fig:high_res}c), aligning with the system’s underlying equations (\ref{eq:fhn_offset}).
The threshold \(\xi_{\text{thres}}=0.49\) was set slightly below the lowest coefficient in Eq.~(\ref{eq:fhn}), \(b=0.5\), facilitating this reduction in complexity. 
Additionally, the coefficient $c'$, which corresponds to the $u^2$ term in the equation for $u_t$, vanishes at $\Delta u = 1/3$ (see Eq.~(\ref{eq:coeff_offset})). Similarly, the coefficient $d'$ for the $u$ term becomes zero at $\Delta u = -1/3$ and $\Delta u = 1$ (see Eq.~(\ref{eq:coeff_offset})). The top panel of Fig.~\ref{fig:high_res}c displays all regions where coefficients are eliminated from the model. The contour plots in Fig.~\ref{fig:high_res}a also show vertical lines at $\Delta u = -1/3, 1/3, 1$. As complexity reduces, the optimal models increasingly converge around these lines where they intersect the nullclines.\\

\begin{figure*}
	\includegraphics[width=\linewidth]{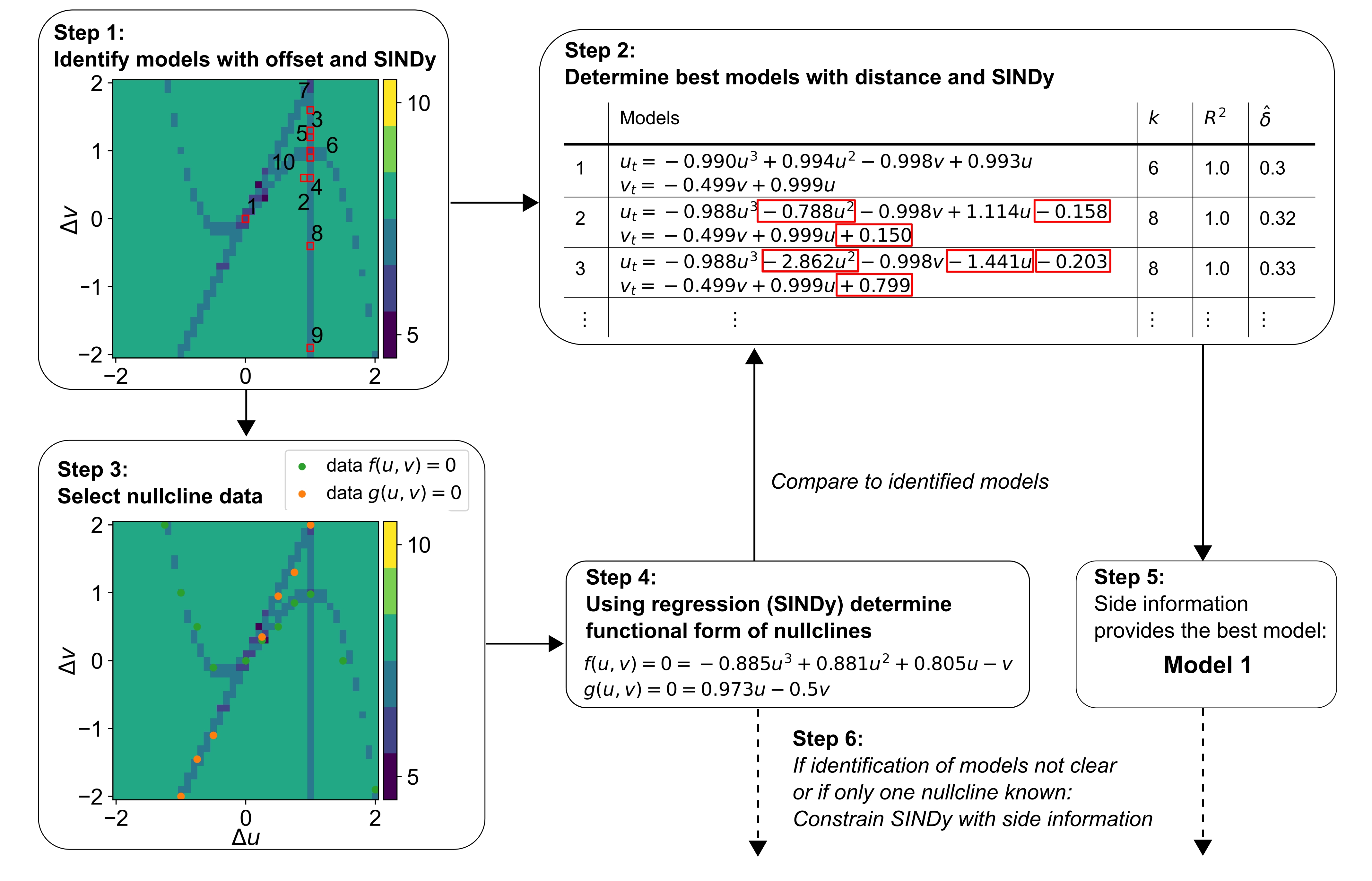}
	\caption{\label{fig:fhn_procedure}  \textbf{Procedure to extract and use additional information to improve model identification using SINDy} For the FHN model with high-resolution data ($dt=0.01$), we identify a subset of effective models. By analyzing the visual patterns of the nullclines, we determine their functional forms, enabling the selection of an optimal model that aligns with the ground truth [Eq.~(\ref{eq:fhn})]. Subsequently, we refine the SINDy model by constraining the term library with this additional information.}
\end{figure*}
\paragraph{Choosing the threshold for optimal nullcline and model identification.}

The nullcline structures and additional vertical lines shown in Fig.~\ref{fig:high_res} depend on the threshold $\xi_{\text{thres}}$. By adjusting this threshold, complexity is controlled by setting some coefficients to zero. We investigated how varying $\xi_{\text{thres}}$ influences model identification and our interpretation of nullcline structures (see Fig.~\ref{fig:fhn_thresh_temp}a,b).\\

Varying the offset and $\xi_{\text{thres}}$ with a fixed temporal resolution ($dt=0.01$) affects model accuracy. Keeping $\xi_{\text{thres}}$ below the smallest coefficient $\xi_{\text{min}}$ results in models that accurately fit the data (see Fig.~\ref{fig:fhn_thresh_temp}a). Exceeding this threshold increases model complexity, the $\hat{R}^2$ score, and the discrepancy measure $\hat{\delta}$, reducing visual information about the nullclines in contour plots. Lowering the threshold provides more detail on the nullclines but at the expense of model performance (higher $\delta_{\text{simp}}$). Optimally, setting $\xi_{\text{thres}}$ slightly below the minimum coefficient ($b = 0.5$) balances model accuracy with nullcline visibility. Alternatively, without prior knowledge of system parameters, aggregating contour plots to focus where $R^2$ scores are highest can identify optimal models (see Fig.~\ref{fig:fhn_thresh_temp}b, last panel).\\

\paragraph{Sufficient temporal resolution is required to extract visual information about the nullclines.}

We varied the temporal resolution $dt$ in our analysis, using the aggregated method to evaluate only the inferred models with the highest $R^2$ scores (see Fig.~\ref{fig:fhn_thresh_temp}c). By subsampling the original data, we increased $dt$ to lower the resolution. As $dt$ increased beyond approximately 0.5, the complexity of the inferred models $\hat{k}$ increased, while their $\hat{R}^2$ scores remained low (see Fig.~\ref{fig:fhn_thresh_temp}c). Further reducing the resolution caused the models to fail in reproducing the initial data and obscured the visual indications of fixed points and nullcline structure (see Fig.~\ref{fig:fhn_thresh_temp}d). However, our findings indicate that extracting additional information remains feasible with temporal resolutions up to $dt=0.1$.

\subsection{Improving model inference using additional information with SINDy}
\begin{figure*}
\includegraphics[width=0.9\linewidth]{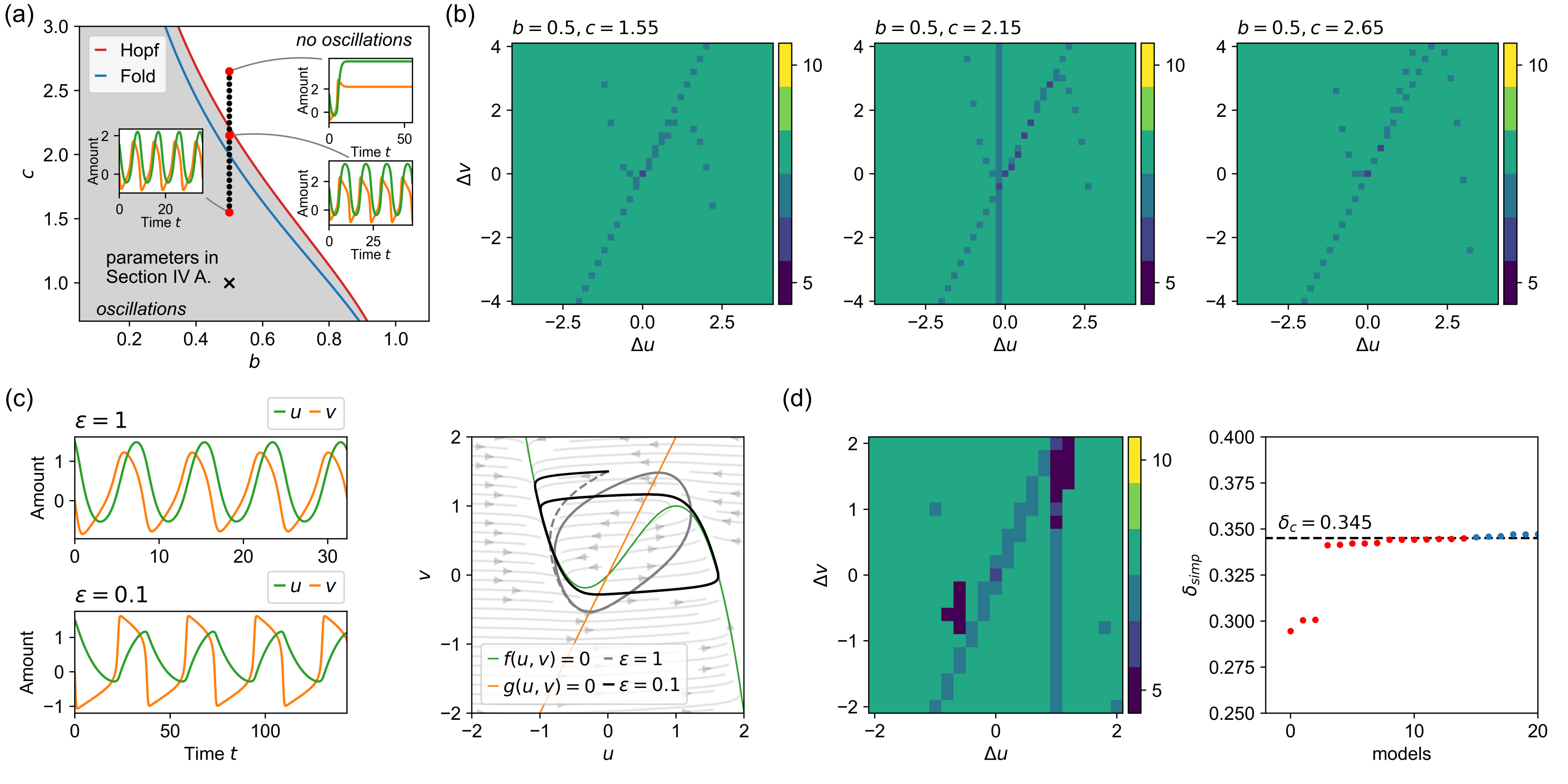}
\caption{\label{fig:bifurcation_timescale} \textbf{Analysis of bifurcation and time scale separation parameters} (a) Varying the parameter $c$ to alter the dynamical regime by crossing the Fold and Hopf bifurcations, along with the corresponding time series of the FHN model. (b) Contour plots of complexity for varying $c$ values show that the nullcline structure can be identified regardless of the distance from the bifurcation, even when the system does not exhibit oscillations. (c) Reducing the parameter $\varepsilon$ enhances the time scale separation, which alters the shape of the limit cycle without changing the phase space landscape. (d) The contour plot of complexity $k$ does not visually reveal the cubic nullcline for $\varepsilon = 0.1$. Nevertheless, the linear nullcline and the correct position of the fixed point can still be identified, even from transient dynamics.}
\end{figure*}
We have demonstrated that by varying the position of a measured attractor (here a limit cycle) in phase space, we are able to visually identify valuable additional information on dynamical properties of the system, i.e. the position of the unstable fixed point and even the nullcline structure. This section outlines a methodology to first extract this information and to subsequently use it to enhance model identification with SINDy (see Fig.~\ref{fig:fhn_procedure}). \\

\textbf{Step 1. Apply SINDy to data with offset to identify models.}
Following the methodology of varying data offsets, we apply SINDy to each dataset. We adjust the threshold $\xi_{\text{thres}}$ across datasets to produce an aggregated contour plot as shown in Fig.~\ref{fig:fhn_thresh_temp}b.\\

\textbf{Step 2. Select optimal models.}
We use the hyperparameter $\delta_{\text{simp}}$ to select the top N models (e.g., $N=15$). The equations for the three highest-performing models are given in Fig.~\ref{fig:fhn_procedure}, highlighting models with 6 or 8 terms. Models 1 and 3, both with 6 terms, perform equally well, although it remains uncertain which corresponds to the actual system (ground truth). \\

\textbf{Step 3. Extract fixed point and/or nullcline equations as additional information.}
The contour plots (Fig.~\ref{fig:fhn_thresh_temp}b) aid in visually pinpointing the fixed point and nullcline structure. We manually select points from each nullcline and use any suitable regression algorithm to extract a functional form. Here, we applied SINDy to approximate their equations:
\begin{align}
	\begin{split}
		\label{eq:SINDy_nullcline}
		 f(\Delta u,\Delta v) &= -0.885 \Delta u^3 + 0.881 \Delta u^2 + 0.805 \Delta u - \Delta v = 0,\\ 
		 g(\Delta u,\Delta v) &=  0.973 \Delta u - 0.5 \Delta v = 0,
	\end{split}
\end{align}
also identifying the unstable fixed point at $(\Delta u, \Delta v) = (0,0)$.\\

\textbf{Step 4. Use additional information to select the best model.}
We compare the nullcline equations with the identified models. Only model 1 matches the structural form, unlike models 2 and 3, which either include additional terms or omit necessary ones.\\

\textbf{Step 5. Refining the model with constrained SINDy.}
Following the selection of Model 1, which aligns with the ground truth, we suggest refining the model by restricting the term library to those in the selected model and reapplying SINDy. This step aims to fine-tune coefficient detection. If the data quality is low, partial information, such as one nullcline or the unstable fixed point alone, may still refine model constraints using constrained SR3 optimization\cite{Champion2019sr3} (see Fig.~\ref{fig:fhn_procedure}).


\subsection{Influence of bifurcation and time scale separation parameters on model identification}
\begin{figure*}
	\includegraphics[width=1\linewidth]{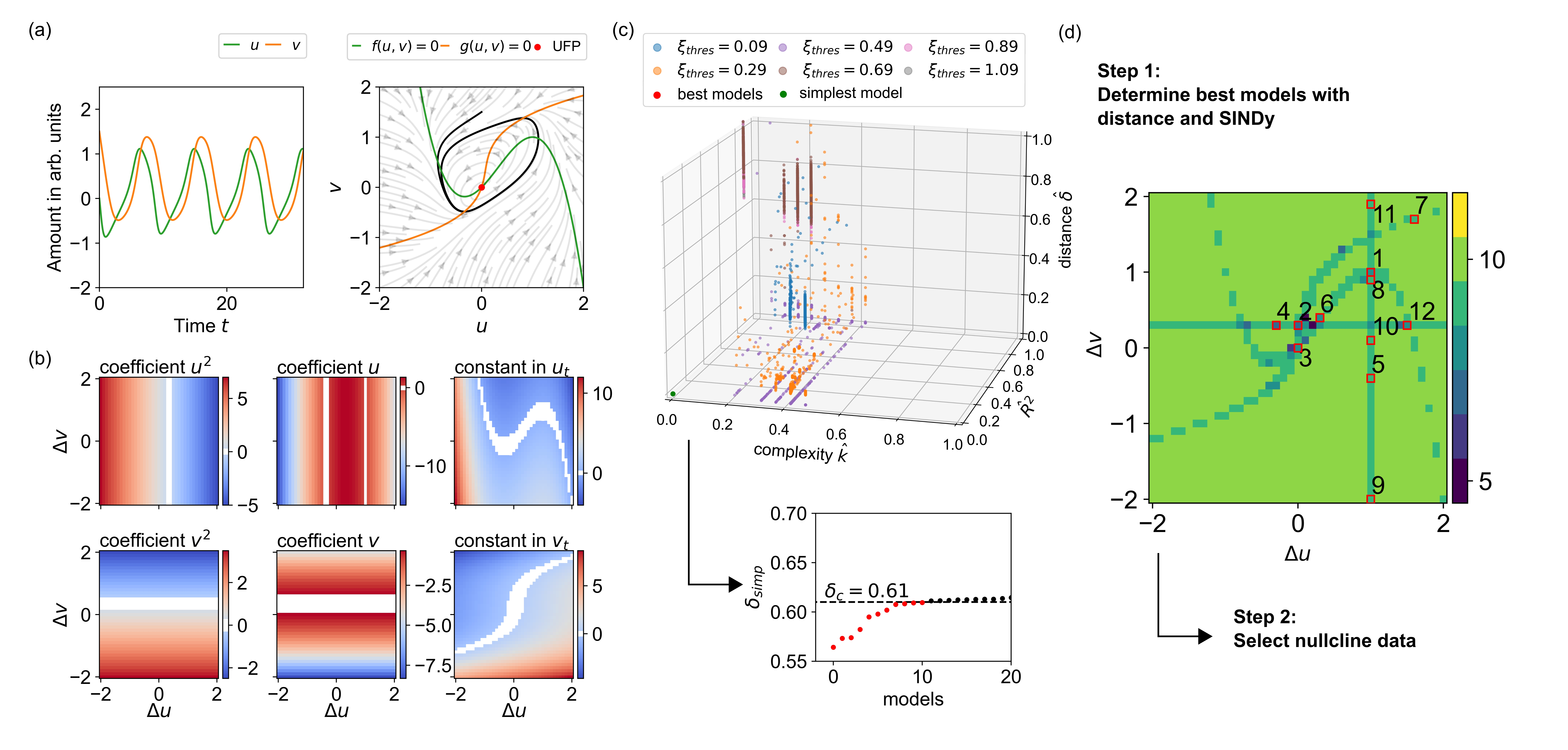}
	\caption{\label{fig:bcb} \textbf{Analysis of bicubic oscillator model} (a) Simulated time series of the bicubic model in Eq.~(\ref{eq:bicubic}) and its representation in the phase space.(b) Values of the identified coefficients for the different offset when a threshold $\xi_{\text{thresh}}=0.29$ is applied. (c) Results of the application of different thresholds in the three-dimensional solution space.(d) The aggregated contour plot of complexity $k$ allows to determine the best models using SINDy and optimal distance $\delta_{\text{simp}}$, see also Tab.~\ref{tab:eq_bcb}. The contour plot gives visual clues about the nullcline structure and the UFP of the system. ($dt=0.01$)}
\end{figure*}

Our procedure enables the identification of additional information that can improve the model identification process of the FHN model. We selected parameters far away from the Hopf bifurcation ($c=1$), where oscillations arise (see Fig.~\ref{fig:bifurcation_timescale}a), and used weak time scale separation ($\varepsilon=1$), which significantly affects the identification of oscillatory systems \cite{Prokop2024, Champion2019} (see Fig.~\ref{fig:bifurcation_timescale}c). Therefore we explored how variations in parameters $c$ and $\varepsilon$ affect the extraction of additional information from time series data.

We adjusted the parameter $c$ from 1.55 to 2.65 in increments of 0.05 to cross the Hopf bifurcation. The compiled contour plots for various $c$ values are displayed in Fig.~\ref{fig:bifurcation_timescale}b. These plots reveal the nullcline structure regardless of proximity to the bifurcation point. Remarkably, even transient time series approaching a stable fixed point (no oscillations, $c>2.2$) allows to recover the underlying nullcline structure.

To examine the effect of time scale separation, we set $\varepsilon$ to 0.1, maintaining similar conditions to Section \hyperref[sec:results]{IV A}. This change alters the oscillation period and the limit cycle shape, as seen in Fig.~\ref{fig:bifurcation_timescale}c. Testing various thresholds $\xi_{\text{thres}}$ with a constant temporal resolution of $dt=0.01$ yields consistent results with those observed at weaker time scale separations (see Fig.~\ref{fig:fhn_thresh_temp}a,b). However, increasing the threshold also increases the distance to the simplest model $\delta_{\text{simp}}$ when surpassing the smallest parameter $-\varepsilon b = -0.05$ of the FHN model in Eq.~(\ref{eq:fhn}).Due to the time scale separation, the aggregated complexity contour plot shown in Fig.~\ref{fig:bifurcation_timescale}d does not visually reveal the cubic nullcline. This is explained by the uneven data distribution across slow and fast oscillation phases, which diminishes insights into the $v$ variable on the limit cycle. Nonetheless, we can still identify the linear nullcline and the correct fixed point.


\subsection{An example of higher complexity ~--~ the bicubic model}
\begin{table}
\caption{\label{tab:eq_bcb} Identified models of the bicubic model with varied offset using SINDy shown in Fig.~\ref{fig:bcb}d (with rounded criteria).}
\begin{ruledtabular}
    \begin{tabular}{llccc}
        & Models & $k$ & $R^2$ & $\hat{\delta}$ \\
        \hline
        \textbf{1} & $u_t = -0.995  u^3-1.992  u^2 -0.998  v$ & 8 & 1.0 & 0.16 \\
          & $v_t = -0.495  v^3-0.991  v^2 -0.831  v + 0.999  u + 0.665   $ & & &\\
        \textbf{2} & $u_t = -1.091  u^3+ 1.365  u  -0.996  v  +0.394  $ & 10 & 1.0 & 0.18 \\
          & $v_t = -0.495  v^3+ 0.494  v^2-0.333  v+ 0.999  u + 0.300 $ &  & &\\
        \textbf{3} & $u_t = -0.990u^3+0.994u^2-0.998v+0.993u$ & 8 & 1.0 & 0.17 \\
            & $v_t = -0.497v^3+0.497v^2-0.334v+0.999u$ &  & &\\
        & $\dots$ &&& 
    \end{tabular}
\end{ruledtabular}
\end{table}
We have shown that by varying the threshold and offset applied to the provided time series data from the FHN model enables the extraction of additional information that enhances model identification. To assess the applicability of our method to more complex scenarios, we explored its effectiveness on a bicubic model with quadratic and cubic terms in the equation for $v_t$: 
\begin{align}
\begin{split}
\label{eq:bicubic}
    u_t &= -u^3 + u^2 + u -v, \\
    v_t &= -\frac{1}{2}v^3 +\frac{1}{2}v^2 - \frac{1}{3} v +u.
\end{split}
\end{align}
\begin{figure*}
\includegraphics[width=0.9\linewidth]{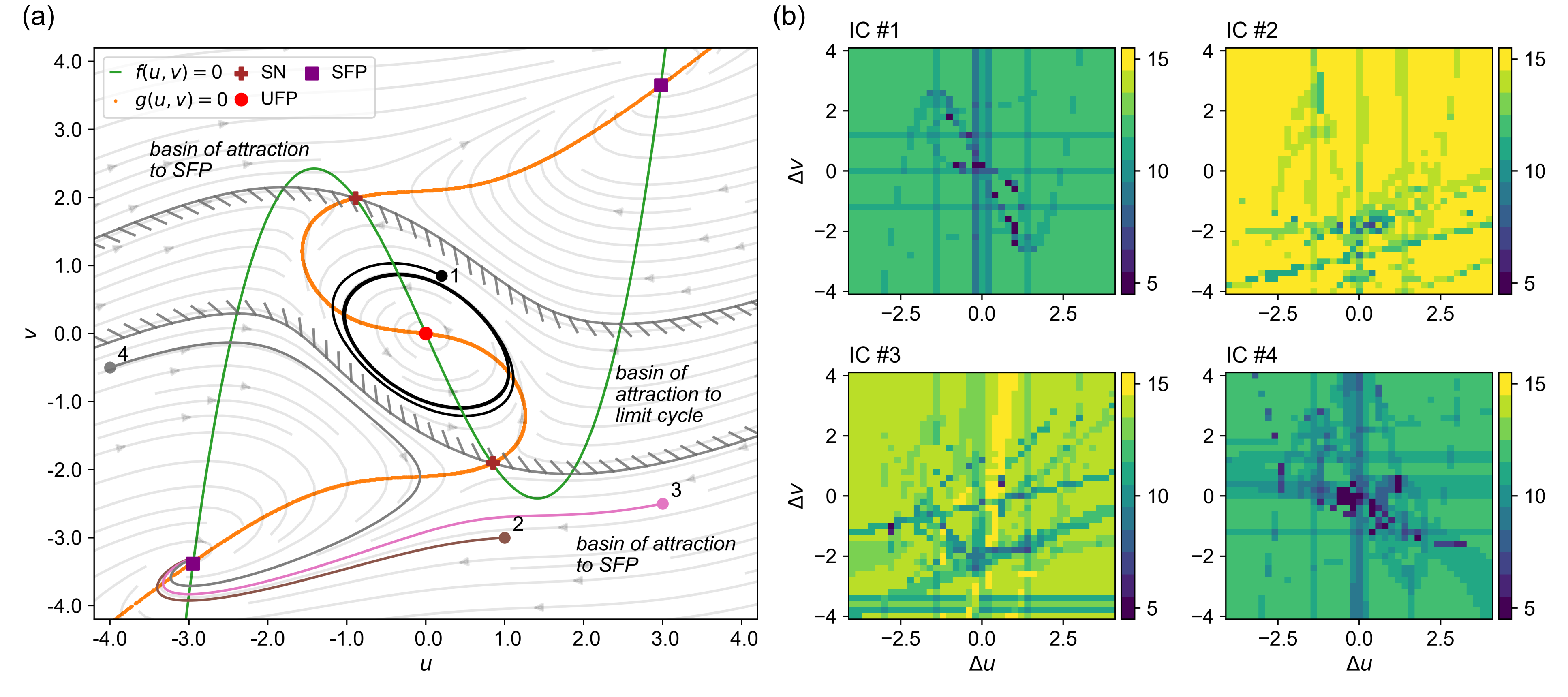}
\caption{\label{fig:glycol} \textbf{Analysis of a multistable model of glycolytic oscillations}
(a) Phase space representation of the glycolytic oscillator model (Eq.~(\ref{eq:glycol_equation})) featuring various dynamic elements: a stable limit cycle, two stable fixed points (SFP), an unstable fixed point (UFP), and two saddle nodes (SN).  
(b) Outcomes of applying our procedure to different initial conditions (ICs) shown in (a). IC \#1 is positioned within the basin of attraction of the limit cycle, while ICs \#2-4 converge to a stable fixed point (SFP). The nullcline structure is clearly detected from long-term oscillatory time series data (IC \#1), but not from short transient dynamics towards a stable fixed point (ICs\#2-4).}
\end{figure*}

We adjusted this model's parameters to align the resulting time series with those from the FHN model (\ref{eq:fhn}), shown in Fig.~\ref{fig:bcb}a, to test if the bicubic model can still be correctly identified when its time series are visually indistinguishable from those in the FHN model (Fig.~\ref{fig:fhn_ts}a). Both nullclines of this bicubic model, depicted in Fig.~\ref{fig:bcb}a, are now S-shaped as experimentally measured in the early embryonic cell cycle \cite{Pomerening2003, Sha2003,Mochida2016,Rata2018,Kamenz2021}.
Recent theoretical studies have analyzed that such oscillations driven by interlinked S-shaped nullclines could yield more robust oscillations \cite{DeBoeck2021,Parra-Rivas2023}.\\


Our procedure, as outlined in Fig.~\ref{fig:fhn_procedure}, begins by adjusting the offset and threshold (here $\xi_{\text{thresh}}=0.29$), which can simplify the model by eliminating certain terms and thus clarifying the structure in the contour plots (Fig.~\ref{fig:bcb}b). Varying the threshold enabled us to select the ten most accurate models (Fig.~\ref{fig:bcb}c), listed in Tab.~\ref{tab:eq_bcb}. These results suggest that the ground truth model (\ref{eq:bicubic}) is not always the top match. By aggregating the threshold contour plots of complexity $k$ (Fig.~\ref{fig:bcb}d), we could deduce the complex cubic nature of both nullclines and derive their equations through regression:

\begin{align}
\begin{split}
\label{eq:SINDy_nullcline_bcb}
f(\Delta u,\Delta v) &= -0.885 \Delta u^3 + 0.881 \Delta u^2 + 0.805 \Delta u - \Delta v,\\ 
g(\Delta u,\Delta v) &= -0.571 \Delta v^3 + 0.567 \Delta v^2 - 0.311 \Delta v + \Delta u. 
\end{split}
\end{align}

Comparing the functional form of the detected nullclines to the ten identified best models in Tab.~\ref{tab:eq_bcb}, we see that only model 3 has the same terms and shares the correct position of the UFP at $(u,v)=(0,0)$. Our approach is thus also able to extract a model equation with the correct form, and with coefficients that are close to the ground truth. Upon reapplying SINDy with a limited set of identified terms, we then accurately captured the model coefficients, affirming the utility of our approach in extracting complex model dynamics accurately.

\subsection{Identification with multiple basins of attraction}

We have studied oscillatory systems with a single basin of attraction that inevitably lead to limit cycle oscillations regardless of initial conditions. This prompted us to question whether it is possible to discover nullcline structures in a multistable system without data from the stable limit cycle attractor.

To explore this, we used a rescaled model of glycolytic oscillations previously introduced \cite{Prokop2024}:
\begin{align}
\label{eq:glycol_equation}
    \begin{split}
        u_t&= au+bv+cu^2+du^3+ev^3, \\
        v_t&= fu+gv+hu^3, \\ 
    \end{split}
\end{align}
with parameters $a=-0.3, b=-2.2, c=-0.25, d=-0.5, e=0.5, f=1.8, g=0.7, h=-0.3$.
As illustrated in Fig.~\ref{fig:glycol}a, the model features three stable solutions: two stable fixed points and one stable limit cycle. 
The basins of attraction of these solution are separated by the stable manifolds from two saddle nodes. 
We used time series data from four distinct initial conditions—one producing oscillations and three inducing varying transient dynamics towards a stable steady state.

Figure~\ref{fig:glycol}b displays the aggregated contour plots of complexity \(k\), omitting the lowest threshold \(\xi_{\text{thres}}\) as detailed in Appendix~\ref{app:lowest_threshold}. This method enabled us to visually reveal the nullcline structure for initial conditions within the basin of attraction of the limit cycle (IC \#1). 
While initial conditions leading to stable fixed points did not generally reveal clear nullcline structures (ICs \#2 and \#3), IC \#4 did, due to its broader exploration of the phase space. 

\section{Discussion}

In this work, we demonstrated that by varying an offset to the time series of an oscillatory system and applying the SINDy white-box model identification method, alongside evaluation metrics such as complexity $k$, $R^2$ score, and distance to the attractor $\delta$, can reveal the system's nullcline structure and fixed points. This additional information enabled us to accurately select the correct model from various candidates. Our methodology was validated using the Fitzhugh-Nagumo model, characterized by a single cubic equation, and tested for robustness to time scale separation and distance from a Hopf bifurcation. We also successfully applied our method to more complex systems, including a bicubic model and a more complex multistable glycolytic oscillation model.

This approach could be particularly valuable for identifying models from oscillatory data where the system details are unknown. It strategically positions the data in phase space to simplify the model, enhancing interpretability and mitigating offset impacts. With high-resolution data, our approach further constrains the identification process, reducing computational demands and yielding meaningful models.

However, our method's effectiveness in the absence of noise, a known challenge for SINDy \cite{Brunton2016, Schaeffer2017, Thaler2019, Prokop2024}, suggests areas for future research. Exploring noise impacts and management strategies, such as filtering \cite{Lejarza2022, Delahunt2022, Cortiella2023, Prokop2024} or using more robust data collection methods like bootstrapping or multiple trajectories \cite{Fasel2022, Ermolaev2022, Prokop2024}, could further refine our approach. Additionally, integrating other advanced model identification techniques like Symbolic Regression \cite{Schmidt2009} or Symbolic Deep Learning \cite{Cranmer2019}, and validating them with experimental data, could extend our findings.

Our method offers a practical solution for using SINDy to uncover important additional information about oscillatory systems directly from measured data, effectively narrowing the space of potential models by imposing such constraints.

\begin{acknowledgments}
L.G. acknowledges funding by the KU Leuven Research Fund (grant number C14/23/130) and the Research-Foundation Flanders (FWO, grant number G074321N).
\end{acknowledgments}

\section*{Conflict of Interest Statement}

The authors have no conflicts to disclose.

\section*{Data Availability Statement}

The data that support the findings of this study are openly available in \textsc{Gitlab} \cite{Gitlab2024} and as an archived repository on RADAR by KU Leuven \cite{XX}.

\appendix

\section{Model identification criteria}
\label{app:criterion}
In our analysis, we evaluate three criteria: complexity $k$, the $R^2$ score, and the distance $\delta$ to the attractor. Complexity refers to the count of non-zero terms in the models. The $R^2$ score, a standard measure of fit, quantifies how well models predict dynamical behavior \cite{Draper2014}. It is defined as:

\begin{equation}
    R^2 = 1 - \frac{\sum_i \left( X_i - N\left[u, v, uv, u^2, v^2,\dots, \mathbf{\xi}\right]\right)^2}{\sum_i \left( X_i - \bar{X}\right)^2 },
\end{equation}
where $\mathbf{X}$ represents measured data and $\bar{X}$ its mean. An $R^2$ score of 1 indicates perfect prediction, while a score of 0 indicates no predictive capability, and a negative score indicates a model that fails to describe any behavior in the data.\\

The distance to the attractor $\delta$ assesses whether a model can reproduce the original limit cycle after perturbation. To compute $\delta$, we simulate the model under 50 varied initial conditions $\mathbf{Y}_{\text{per}}=(w,z)$ and measure the average, relative distance to the original limit cycle data $\bm{X}=(u,v)$:

\begin{equation}
    \delta = \frac{\sum_i \min \sqrt{(u_i - w_j)^2+(v_i-z_j)^2}}{\sum_i (u_i - \bar{u})^2+(v_i - \bar{v})^2}.
\end{equation}

When the distance $\delta$ approaches 0 then the model reproduces the same limit cycle despite the perturbation thus being more general. 

\section{Discarding lower thresholds for analysis}

\begin{figure}
\includegraphics[width=0.8\linewidth]{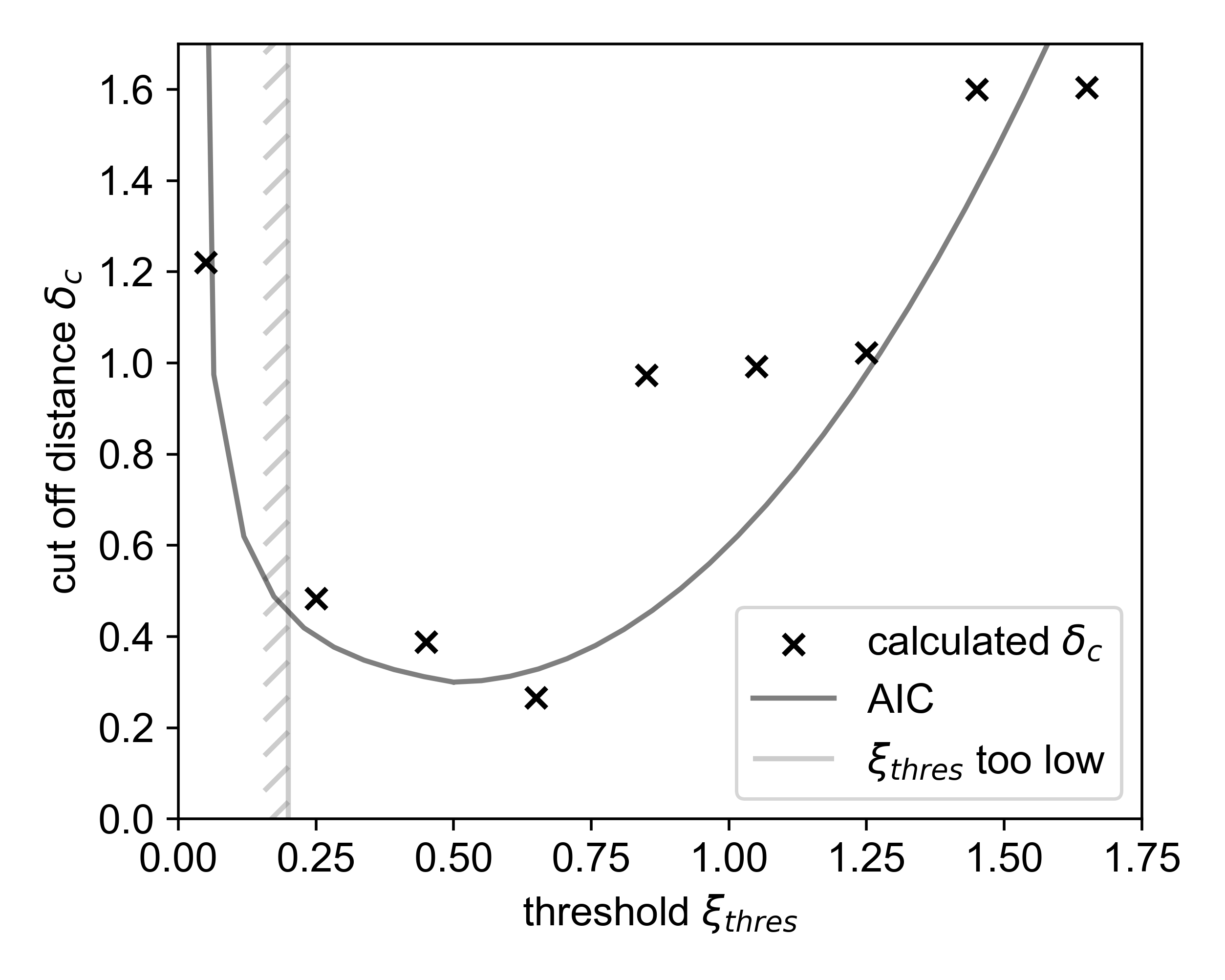}
\caption{\label{fig:threshold} Lower evaluated thresholds result in significant overfitting, depicted as a large distance to the simplest model ($\delta_{\text{simp}}$) and corresponding cutoff distance ($\delta_c$), which behave similarly to the Akaike Information Criterion (AIC)}
\end{figure}
\label{app:lowest_threshold}

To generate aggregated contour plots of complexity $k$ for visual insights into phase space structure, we typically select the threshold $\xi_{\text{thres}}$ where the R$^2$ score is highest for each evaluated offset $(\Delta u, \Delta v)$. This selection effectively prevents overfitting and unnecessary model terms in simpler models like the FitzHugh-Nagumo and bicubic models. However, for complex models such as the glycolytic oscillator, adopting all thresholds ($\xi_{\text{thres}}=[0.05, 1.65]$ in $0.2$ steps) can yield poor outcomes. At the lowest threshold, overfitting occurs, characterized by the selection of numerous additional terms that fail to generalize (indicated by a large $\hat{\delta}$ and $\delta_{\text{simp}}$, as shown in Fig.~\ref{fig:threshold}), resulting in inflated R$^2$ scores. The change of $\delta_{\text{simp}}$ in dependance of $\xi_{\text{thres}}$ resemble the change of Akaike Information Criterion \cite{Akaike1973} (shown conceptually in Fig.~\ref{fig:threshold}), where too low thresholds provide candidate models are too complex and are penalized, while too high thresholds result in low-complexity models which do not provide sufficient information about the undelying system.

\bibliography{bibliography.bib}

\end{document}